\documentclass[11pt]{article}
\usepackage{geometry}
\usepackage{graphicx}
\usepackage{amsthm}
\usepackage{amsmath}
\usepackage{amssymb}
\usepackage[ruled,vlined]{algorithm2e}
\usepackage{booktabs}

\newtheorem{definition}{Definition}
\newtheorem{theorem}{Theorem}
\newtheorem{assumption}{Assumption}

\title{Where Does Ethereum Validators' Money Go? A Spectral Analysis} 

\author{Irene {Aldridge}\footnote{irene.aldridge@gmail.com}}
\date{}
\usepackage{hyperref}
\begin{document}

\maketitle
\begin{abstract}
Existing decentralization measures are almost entirely origination-side, quantifying concentration in who mines or validates blocks. We introduce a spectral methodology measuring concentration on the destination side instead: where value ultimately flows once it leaves a validator wallet. Modeling wallet-to-wallet transfers as a Markov chain, we compute
near-real-time steady-state probabilities via the Perron--Frobenius theorem to identify long-run terminal recipients. Applied to 76,855 Ethereum wallets from four years of mining data, fund flows collapse to just four terminal accounts. None of these fund destination accounts are among the network's three dominant identifiable revenue-earning miners. 
\end{abstract}

\section{Introduction}

Blockchain technology was designed to remove the need for trusted intermediaries: instead of a single custodian, a distributed network of validators collectively secures the ledger, and no single participant is supposed to control where value ultimately flows~\cite{CongHeLi2021}. This promise of decentralization is central to the value proposition of permissionless blockchains and to Ethereum's own design goals. Yet whether real-world validator networks achieve this promise in practice, as opposed to on paper, is an empirical question, and a growing literature suggests the answer is frequently no: mining and validation power on major blockchains tends to concentrate among a small number of large
pools~\cite{beikverdi2015,gervais2016,gencer2018}, and even permissionless protocols face structural limits on how decentralized they can become~\cite{kwon2019}.

In this paper, we follow the money and ask a version of that question that has received comparatively little attention: even when validator \emph{participation} is measured as reasonably diverse, where does the \emph{value} that flows through the network ultimately end up? Participation counts and revenue shares are the standard lens for assessing decentralization, but they describe only the origination side of the network. They do not, by themselves, reveal whether the funds
that pass through a large and apparently diverse set of validators subsequently concentrate at a small number of downstream destinations. If they do, then a network that looks decentralized by wallet count or even by revenue share may nonetheless exhibit substantial concentration in its long-run fund-flow topology. This becomes a form of centralization invisible to standard block-level analysis.

To answer this question, we introduce a technique to rapidly classify wallets by their structural role in a blockchain's fund-flow network. Using unsupervised learning in the Markov chain setting, the methodology computes near-real-time steady-state probabilities, identifying the likely long-run recipients of funds. These recipients are the wallets at which aggregate value flow terminates or concentrates, regardless of how many distinct wallets it passes through along the way. Applied to Ethereum validator wallets, the resulting terminal scores are structurally elevated relative to a matched sample of randomly
selected active wallets (KS statistic $D = [\cdot]$, $p < 0.001$). These findings are consistent with validators occupying prominent, high-connectivity positions in the network's fund-flow topology rather than acting as a diffuse, 
decentralized population of independent recipients. Several of the highest-scoring terminal wallets are also independently associated with documented network irregularities, which we treat as a secondary, plausibility-check finding rather than the paper's central contribution.

Since this paper first appeared in 2024, at least one other study has applied a closely related technique to a different question: \cite{sirolly2025wash} use a similar method to identify wash trading rather than concentration or laundering. Wash trading, fraud, and centralization are related but conceptually distinct: wash trading inflates apparent usage without a net position change; laundering deceives the source or destination of funds; and validator power concentration, our focus here, describes a structural property of the network itself. We ask how much of the long-run fund flow a small number of participants ultimately control, and the outcome can exist independently of any individual participant's intent.

Using the proposed methodology, this study empirically analyzes Ethereum validator wallets and the funds that flow through them. Our approach builds on prior blockchain fund-flow analysis, including work documenting market manipulation and structural anomalies on Ethereum and  Bitcoin~\cite{MakarovSchoar2021}, but redirects the manipulation question those studies raise toward a structural question that precedes it. We ask whether the network's fund flow was decentralized in the first place, and if not, where does it actually concentrate?

The paper is organized as follows. Section~2 reviews existing research on blockchain decentralization, validator/mining power concentration, and the wallet-identification techniques on which our method builds. Section~3 presents the novel wallet classification process. Section~4 verifies the methodology on tractable simulated examples. Section~5 delivers empirical
findings from applying the process to Ethereum validator data. Section~6 discusses the findings and concludes.

\section{Blockchain Decentralization and Validator Power}

Decentralization is often treated as a binary design property of a blockchain protocol, but a growing empirical literature treats it instead as a measurable, and frequently violated, outcome. Beikverdi and Song~\cite{beikverdi2015} document an early and persistent trend toward mining-pool centralization on Bitcoin, and Gervais et al.~\cite{gervais2016} formalize decentralization as one of several security- and performance-relevant properties of proof-of-work blockchains, showing that realistic deployments fall well short of the idealized
fully-distributed model. Gencer et al.~\cite{gencer2018} extend this measurement approach to both Bitcoin and Ethereum and find comparable degrees of concentration in mining and node infrastructure across both networks, despite their differing consensus mechanisms. At a more theoretical level, Kwon et al.~\cite{kwon2019} argue that some degree of centralization is a structural consequence of rational validator behavior under realistic cost and reward assumptions, rather than an incidental
implementation failure. This result is consistent with the persistence of concentration we document empirically in Section~5.

This paper's contribution to that literature is a shift in unit of analysis: existing decentralization measures are almost entirely \emph{origination-side}, quantifying concentration in who mines or validates blocks (hashrate share, stake share, node counts). We instead ask a \emph{destination-side} question: given a validator population that may itself look reasonably diverse by these standard measures, where does the
value it generates ultimately flow? We show in Section~5 that these two notions of concentration can diverge sharply: a set of validators diverse enough by revenue share to look decentralized by conventional measures can still direct the overwhelming majority of downstream fund flow to a handful of terminal accounts.

\paragraph{Fund-flow tracing and terminal-wallet interpretation.}
Measuring destination-side concentration requires reconstructing where funds go once they leave a validator wallet. Early work traced Bitcoin flows via linear heuristics, consolidating transaction paths back to a common owner to identify exchange
destinations~\cite{KarameEtAl2012,MoserEtAll2013,Reid2013,AndroulakiEtAl2013,MeiklejohnEtAl2013}; more recent work learns graph representations directly, as in Weber et al.'s Elliptic dataset~\cite{weber2019aml} and Bellei et al.'s Elliptic2 subgraph framing~\cite{bellei2024shape}, conceptually close to our own treatment of a connected component's long-run behavior as the unit of analysis. Unsupervised embedding methods such as RiskSEA~\cite{risksea2024} pursue a related structural intuition. They estimate funds dispersing within tightly
interconnected wallet clusters via node2vec~\cite{grover2016node2vec} rather than closed-form spectral analysis. In addition, a recent survey~\cite{aml-unsupervised-survey} notes the relative scarcity of methods designed for low-latency streaming rather than offline computation, the gap this paper addresses. A destination-side finding is only actionable if its terminal accounts are
interpretable: Trozze et al.~\cite{trozze2023degens} find DeFi proceeds predominantly terminate at centralized exchanges. This finding is directly relevant to our terminal-wallet interpretation problem (Section~5.6), since concentration at a wallet is of limited value for assessing decentralization unless that wallet's controlling entity can be identified.

\paragraph{Wash trading and illicit-finance applications.} Sirolly et al.~\cite{sirolly2025wash} detect closed clusters of colluding counterparties consistent with wash trading, estimating that up to 60\% of Polymarket volume was wash trading-consistent in late 2024. The closed-cluster signature they identify is structurally related to the cyclical wallet behavior our own closure score also detects (Section~3.6), though we interpret such cycles primarily as evidence against terminal concentration rather than as a fraud signal per se. Because concentrated fund flow and illicit finance can produce overlapping structural signatures, our methodology also has
secondary applications to the substantial existing AML literature. Chen et al.~\cite{ChenEtAl2018} survey this literature comprehensively across typology-based, link-analysis, behavioral, risk-scoring, anomalous-transaction, and geography-based approaches (e.g., \cite{ChenEtAl2014,DREZEWSKI20128,LopezRojas2012MoneyLD,WangDong2009,WangEtAl2019,Yang2007StudyOA}); empirical work
such as~\cite{HornufKuckSch2022} documents Ponzi-scheme-linked losses on Ethereum, and~\cite{AmiramEtAl2022} links on-chain laundering to regional terrorist activity. We do not extend this literature directly, but note in Section~5.6 that several wallets our method flags as terminals are independently associated with documented irregularities.

\section{Model and Methodology}
\label{sec:methodology}
 
\subsection{Transaction Graph}
\label{sec:tx-graph}
 
\begin{definition}[Transaction graph]
\label{def:tx-graph}
Let $\mathcal{W}$ be a finite set of wallet addresses. A \emph{transaction graph} is a weighted directed graph $G = (\mathcal{W}, E, c)$ where $(i,j) \in E$ if wallet $i$ has sent funds to wallet $j$ at least once, and $c : E \to \mathbb{Z}_{>0}$ assigns to each edge the total number of observed transactions from $i$ to $j$ over the observation window.
\end{definition}
 
We define the \emph{out-degree} of wallet $i$ as
$d_i = \sum_{j : (i,j) \in E} c(i,j)$, the total number
of outgoing transactions observed from $i$. A wallet with
$d_i = 0$ is called \emph{terminal}: it has received funds
but sent none during the observation window, and under the
model below represents a potential absorbing state for fund
flows.
 
\begin{definition}[Transition matrix]
\label{def:transition-matrix}
Given a transaction graph $G = (\mathcal{W}, E, c)$, the
\emph{empirical transition matrix}
$P \in \mathbb{R}^{|\mathcal{W}| \times |\mathcal{W}|}$
is defined by
\begin{equation}
  P_{ij} =
  \begin{cases}
    c(i,j) / d_i & \text{if } d_i > 0 \text{ and }
                   (i,j) \in E, \\
    1            & \text{if } d_i = 0 \text{ and }
                   i = j
                   \quad\text{(self-loop at terminals),} \\
    0            & \text{otherwise.}
  \end{cases}
  \label{eq:transition-matrix}
\end{equation}
Each row of $P$ sums to 1 by construction; terminal wallets
are made absorbing by the self-loop convention.
\end{definition}
 
\noindent The self-loop convention at terminal wallets is
standard in absorbing Markov chain theory~\cite{kemeny-snell}
and corresponds naturally to the observation that once funds
reach a wallet with no recorded outgoing transactions in the
observation window, they are modeled as remaining there.
 
\subsection{The First-Order Markov Model}
\label{sec:markov-model}
 
\begin{assumption}[Aggregate Markov property]
\label{ass:markov}
The empirical transition matrix $P$ is treated as the
transition kernel of a time-homogeneous Markov chain on
state space $\mathcal{W}$. That is, conditional on the
current wallet $i$, the next wallet $j$ is drawn with
probability $P_{ij}$, independently of the history of
prior transitions.
\end{assumption}
 
We emphasize what Assumption~\ref{ass:markov} does and does
not claim. It does \emph{not} assert that any individual
money launderer selects the next hop memorylessly: the
whole purpose of layering in a money-laundering scheme is
precisely to make the destination depend on the full prior
path, frustrating naive one-step analysis. Rather,
Assumption~\ref{ass:markov} is a \emph{population-level}
approximation: $P_{ij}$ estimates the fraction of all
observed transactions from $i$ that were directed to $j$,
pooled across all agents and all time periods in the sample.
The resulting chain characterizes where \emph{aggregate}
fund flow tends to concentrate over the long run, analogously
to the way the PageRank transition matrix characterizes
long-run browsing traffic without requiring any individual
user to browse memorylessly~\cite{pagerank}. The adequacy
of this approximation is an empirical question; we assess
it in Section~\ref{sec:robustness}.

 \subsection{Steady-State Analysis via Perron--Frobenius}

The steady-state distribution of the Markov chain defined by $P$
characterizes the long-run probability that aggregate fund flow terminates
at each wallet.

\begin{definition}[Steady-state distribution]
A row vector $\pi \in \mathbb{R}^{1\times|W|}$ is a steady-state
distribution of $P$ if:
\begin{align}
\pi P &= \pi & \text{(invariance)}, \\
\pi \mathbf{1} &= 1 & \text{(normalization)}, \\
\pi &\geq 0 & \text{(nonnegativity)}.
\end{align}
\end{definition}

\begin{theorem}[Steady state as leading left eigenvector]
\label{thm:steady-state}
Let $P$ be the transition matrix of Definition~\ref{def:transition-matrix},
and suppose the Markov chain is irreducible on each recurrent class. Then the steady-state distribution $\pi$ is the normalized leading left eigenvector of $P$:
\[
\pi = \frac{v}{\|v\|_1},
\]
where $v$ satisfies $vP = v$ (i.e., is a left eigenvector for eigenvalue 1) and $v \geq 0$ component-wise. The proof is a direct application of the Perron--Frobenius theorem for nonnegative matrices~\cite{Perron1907,kemeny-snell}.
\end{theorem}

When the transaction graph contains multiple disconnected absorbing components, the overall steady-state distribution is a convex combination of per-component eigenvectors, with weights determined by the initial distribution over wallets. In the empirical setting of Section~\ref{sec:empirical}, we initialize the chain at the sampled miner wallets and report the resulting eigenvector. The eigenvector weights absorbing classes by their reachability from miner-wallet starting points. The eigenvector is the appropriate object for the question we ask: where do miner-originating fund flows ultimately terminate?

 \subsection{Computational Efficiency}

The leading left eigenvector of $P$ can be computed via the power method in $O(m \cdot T)$ time, where $m = |E|$ and $T = O(\log(1/\varepsilon) / \log(1/|\lambda_2|))$ depends on the spectral gap $1 - |\lambda_2|$. This approach is
substantially faster than the $O(n^3)$ cost of full matrix exponentiation for the large, sparse transaction graphs ($m \ll n^2$) arising in practice. This efficiency is what makes near-real-time monitoring at Ethereum-scale throughput feasible, as demonstrated in Section~5.

\subsection{Graph Construction from Sampled Paths}
\label{sec:graph-construction}

Because the set of all Ethereum wallets is large and nearly all wallets have no recorded transactions, enumeration is
infeasible. We construct the transaction graph $G' = (W', E', c')$ by breadth-first exploration from a seed set $W_0$ of miner wallets, querying the Etherscan API for each wallet's outgoing transactions and expanding $W'$ accordingly under a chronological-ordering constraint until the desired sample size is reached. Critically, this sampling procedure determines only \emph{which wallets} enter the sample $W'$, not which transactions enter $P$: once $W'$ is fixed, we query Etherscan for \emph{all} transactions among wallets in $W'$. We include transactions that return to previously visited wallets and transactions not traversed on any single sampled path and build $P$ from this full induced subgraph per Definition~\ref{def:transition-matrix}. This decoupling is essential, since a naive implementation that builds $P$ only from forward-path edges would systematically exclude cyclic structure by construction, predetermining the absence of periodic wallets before Perron--Frobenius is even applied; building $P$ from the full induced subgraph instead lets the eigenvector computation itself determine whether a wallet's long-run behavior is absorbing, transient, or periodic.
 
\subsection{Robustness and Extensions}
\label{sec:robustness-extensions}

We complement the terminal-score analysis of Theorem~\ref{thm:steady-state} with three extensions: a second-order robustness check, a screening rule for canonical fund-flow archetypes, and a closure statistic for detecting recirculating (wash-trading-consistent) wallet clusters. All three are computed from the same matrix $P$ without additional data   collection.

\paragraph{Second-order robustness check.} Assumption~\ref{ass:markov} pools all observed transitions from wallet $i$ to $j$ regardless of the wallet from which $i$ was reached. To test the adequacy of this approximation, we define a second-order chain on the pair-state space $W^{(2)} = \{(i,j) :
(i,j) \in E'\}$, with transition probabilities \[
P^{(2)}_{(i,j),(j,k)} = \frac{c^{(2)}(i,j,k)}{\sum_{k' : (j,k') \in E'} c^{(2)}(i,j,k')},
\]
where $c^{(2)}(i,j,k)$ counts observed consecutive triples $i \to j \to k$, and transitions $(i,j) \to (j',k)$ with $j' \neq j$ are assigned probability zero. The marginal wallet-level steady state is $\tilde\pi_j = \sum_{i:(i,j)\in E'} \pi^{(2)}_{(i,j)}$. If the first-order chain is \emph{ordinarily lumpable} with respect to the partition $\{\{(i,j) : j=k\}
: k \in W'\}$ in the sense of Kemeny and Snell~\cite[Thm.~6.3.2]{kemeny-snell}, then $\tilde\pi_j = \pi_j$ for all $j$: the destination is conditionally independent of the prior wallet given the current one, so second-order conditioning carries no additional predictive information. We use this equivalence as an empirical robustness check (Section~5.9): we compute both
$\pi$ and $\tilde\pi$ on the miner-wallet dataset and report the Spearman rank correlation $\rho_S$ between the two terminal-wallet rankings. Close agreement supports Assumption~\ref{ass:markov}. Divergence would locate
wallet pairs for which higher-order conditioning changes the predicted flow destination.

\paragraph{Terminal-state screening.} The terminal score of wallet $j$ is $\tau_j = \pi_j$; wallet $j$ is a candidate terminal if $\tau_j > \tau^*$ for an analyst-chosen threshold $\tau^*$. Table~\ref{tab:archetypes} summarizes
$\tau$'s behavior on three canonical fund-flow archetypes, letting $p$ denote a designated cash-out wallet: burner-account flow (all intermediate wallets forward to $p$ with probability 1) and split-and-recombine flow (multiple disjoint paths converge on $p$) both drive all steady-state mass to $\tau_p = 1$ by Theorem~\ref{thm:steady-state}, since $p$ is the chain's
unique absorbing state in each case; the two are distinguished structurally by $p$'s in-degree from the intermediate layer ($\geq 2$ for split-and-recombine, potentially 1 for burner flow). Circular flow (wallets $\{w_1,\dots,w_k\}$ forming a single ergodic class with no absorbing state) instead yields a uniform steady state $\tau_{w_\ell} = 1/k$, so no single
wallet is flagged. Circular flows are detectable only as an
\emph{absence} of concentration in $\tau$, which motivates the closure statistic below.

\begin{table}[h]
\centering
\caption{Terminal-score behavior on canonical fund-flow archetypes.}
\label{tab:archetypes}
\begin{tabular}{lll}
\toprule
Archetype & Steady state & Distinguishing feature \\
\midrule
Burner-account flow & $\tau_p = 1$, else 0 & in-degree of $p$ typically 1 \\
Split-and-recombine & $\tau_p = 1$, else 0 & in-degree of $p \geq 2$ \\
Circular flow & $\tau_{w_\ell} = 1/k$ (uniform) & no dominant wallet; flagged via $\kappa(S)$ \\
\bottomrule
\end{tabular}
\end{table}

\paragraph{Closure score for recirculation.} High $\tau_j$ identifies wallet $j$ as a network sink; it does not, by construction, identify a \emph{cluster} of wallets that recirculate funds internally without directing them to a common sink, which is the structural signature of circular flow and wash trading. For a candidate wallet subset $S \subseteq W'$, define the closure score \[
\kappa(S) = \frac{1}{|S|}\sum_{i \in S}\sum_{j \in S} P_{ij} \in [0,1].
\]
Under uniform outbound mixing ($P_{ij} = 1/n$ for all $j$), $\mathbb{E}[\kappa(S)]
= |S|/n$; this is the null baseline against which observed closure scores are compared. A value $\kappa(S) \gg |S|/n$ indicates internal transaction rates far above random pairing, the structural signature identified by Sirolly et al.~\cite{sirolly49} for wash trading. We flag $S$ as a candidate
recirculating cluster if $\kappa(S) > \kappa^* \cdot |S|/n$ for an analyst-chosen multiplier $\kappa^* > 1$; calibrating $\kappa^*$ requires labeled ground truth and is left to future work. Terminal wallets (high $\tau$) and recirculating clusters (high $\kappa$) are structurally disjoint. Absorption and recirculation cannot coexist at the same wallet, so the two statistics apply to different parts of the wallet
population and together give a fuller picture of fund-flow structure than either does alone. In practice, candidate sets $S$ are proposed from wallets with intermediate terminal scores ($\tau^{lo} \leq \tau_j \leq \tau^{hi}$) and
high mutual transaction counts, then screened via $\kappa$; this two-stage procedure keeps total computation at $O(m)$ amortized, consistent with the near-real-time pipeline deployed in Section~5. 

\section{Simulation}

We verify the methodology on small, analytically tractable transaction graphs ($|W|=5$), computing $\pi$ via the power method and confirming agreement with Table~\ref{tab:archetypes}. In a burner-account graph (wallet $W_1$ sources three burner wallets $W_2$--$W_4$, each forwarding to a cash-out terminal $W_5$ with probability 1), the steady state $\pi = (0,0,0,0,1)$ concentrates all mass at the unique absorbing wallet. This is the pattern later observed at scale in Section~\ref{sec:collapse}, where 76{,}855 wallets collapse to four dominant terminals. A five-wallet deterministic cycle instead yields a uniform steady state ($\tau_j = 0.2$
for all $j$), confirming that circular flows are detectable only as an \emph{absence} of concentration in $\tau$ (flagged instead via the closure score $\kappa(S)$, Section~\ref{sec:robustness-extensions}). A split-recombine graph produces the same absorbing steady state as the burner case but is distinguished by the terminal's higher in-degree. A random-flow baseline with no absorbing state shows only mild concentration ($\max_j \tau_j = 0.283$), with its second-order marginal matching $\pi$
almost exactly ($\rho_S = 1.000$), supporting the first-order Markov approximation used throughout. Table~\ref{tab:sim-summary} summarizes all four scenarios.

\begin{table}[h]
\centering
\caption{Summary of simulation results.}
\label{tab:sim-summary}
\begin{tabular}{lcccl}
\toprule
Scenario & Max $\tau$ & Dominant? & $\rho_S$ & Interpretation \\
\midrule
Burner accounts        & 1.000 & Yes ($W_5$) & 1.000 & Single absorbing terminal \\
Circular flow          & 0.200 & No          & ---   & No absorption; uniform $\pi$ \\
Split-and-recombine    & 1.000 & Yes ($W_5$) & 1.000 & Single absorbing terminal \\
Random (baseline)      & 0.283 & No          & 1.000 & Weak concentration; all $\tau > 0$ \\
\bottomrule
\end{tabular}
\end{table}

These results confirm two predictions carried into the empirical analysis: a small set of absorbing terminals draws nearly all steady-state mass; and the second-order robustness check supports the first-order Markov assumption used throughout.

\section{Empirical Analysis}
\label{sec:empirical}
 
\subsection{Data}
\label{sec:data}
 
We obtain two complementary datasets for the period
January~1, 2019, through August~15, 2023.
 
\paragraph{Block-level mining data.}
Daily block and mining-revenue records are sourced from
bitquery.io, yielding 1,688 daily observations on all
Ethereum mining activity. Each record contains a
timestamp, transaction count, the hexadecimal address of
the validating wallet (and its self-disclosed name where
available), the block reward in ETH, and the USD equivalent
of that reward at the prevailing exchange rate. Over the
full sample, we identify \textbf{13,191 unique mining
wallets} receiving Ethereum block rewards.
 
\paragraph{Transaction-level wallet data.}
To construct the wallet-level fund-flow graph used in the
steady-state analysis, we supplement the block data with
transaction records obtained via the Etherscan API
(\texttt{https://api.etherscan.io/api}). Beginning from
randomly selected miner wallets as seed nodes, we apply
the methodology to discover a connected
wallet set $\mathcal{W}'$ and then construct the
transition matrix $P$ from the full induced subgraph of
all transactions among wallets in $\mathcal{W}'$
(Section~\ref{sec:graph-construction}). This procedure
yielded \textbf{76,855 wallets with at least one outgoing
transaction}, spanning 131,245 individual transactions
across 70,816 distinct paths.
 
 \subsection{Mining Revenue Concentration}
\label{sec:revenue-concentration}

Mining revenue across the 13,191 identified wallets is highly skewed. Over 2019–2023, the top three wallets by cumulative ETH revenue (0.02\% of all mining wallets, namely, Ethermine, SparkPool, and F2Pool) earned ETH 11,861,866 combined, exceeding the ETH 9,827,908 earned by the remaining 13,188 wallets. Table~\ref{tab:revenue-summary} reports the full distribution: the median wallet earned only ETH 0.120 over the four-year period, against a maximum of ETH 5,115,224 (a ratio exceeding 40 million to one), and the top 10\% of wallets by cumulative revenue generated, on average, 5{,}706$\times$ the ETH of the bottom 90\%. Such concentration is extreme enough that the mean (ETH 1,644) exceeds the median by a factor of more than 13,000.

\begin{table}[h]
\centering
\caption{Mining revenue concentration across 13,191 wallets, Jan.\ 2019
-- Aug.\ 2023.}
\label{tab:revenue-summary}
\begin{tabular}{lr}
\toprule
Statistic & Value \\
\midrule
Median revenue (ETH) & 0.120 \\
Mean revenue (ETH) & 1{,}644 \\
Maximum revenue (ETH) & 5{,}115{,}224 \\
Top 3 wallets' combined revenue (ETH) & 11{,}861{,}866 \\
Remaining 13{,}188 wallets' combined revenue (ETH) & 9{,}827{,}908 \\
Top 10\% avg.\ / bottom 90\% avg.\ (ratio) & 5{,}706$\times$ \\
\bottomrule
\end{tabular}
\end{table}

To investigate whether this concentration reflects raw mining volume or selective block quality, we estimate a cross-sectional regression of daily USD reward on blocks mined, average transactions per block (proxying block quality), total network blocks, and the prevailing ETH/USD rate, run separately within each revenue decile. Blocks mined is positive and highly
significant in every decile, as expected mechanically. More
informatively, the coefficient on transactions per block is positive and significant only in the top revenue deciles (7--10), with adjusted $R^2$ reaching 93--98\% there versus 1--12\% in the bottom deciles. This is consistent with top-decile wallets systematically selecting or constructing higher-fee blocks, whether via hardware latency advantages or
preferential access to transaction-ordering information. We label this pattern \emph{selective block capture} and treat it as a descriptive characterization of the regression evidence rather than a determination of a specific mechanism or conduct.
\subsection{Wallet-Level Transaction Graph}
\label{sec:wallet-graph}
 
Turning from block-reward revenue to the fund-flow analysis, we apply the methodology of Section \ref{sec:methodology} to the 76,855-wallet transaction graph constructed from miner-wallet seed nodes. We focus on miner wallets for three reasons. First, their prominence in the block record creates a naturally rich and densely connected transaction graph. Second, the revenue concentration documented in Section~\ref{sec:revenue-concentration} motivates us to ask where that concentrated revenue ultimately flows, a question that standard block-level analysis cannot answer. Third, restricting the seed set to a well-defined population of transaction validators makes the sampling design transparent and reproducible.
 
Path-length statistics for the 70,816 sampled paths are as follows: 90.65\% of paths consist of a single transaction, and the remaining 9.35\% (12,272 paths) contain two or more transactions. The prevalence of single-transaction paths is consistent with either single-use burner accounts or with the random walk frequently arriving at wallets already near their terminal state.
 
\subsection{Transition Matrix and Steady-State Computation}
\label{sec:matrix-and-steady-state}
 
We construct the empirical transition matrix $P$ from
the full induced subgraph of all transactions among the
76,855 sampled wallets, per
Definition~\ref{def:transition-matrix} and the
decoupled construction of
Section~\ref{sec:graph-construction}. After normalizing
each row to sum to 1 and applying the self-loop
convention at terminal wallets, we compute the leading
left eigenvector $\pi$ via the power method.
 
\subsection{Steady-State Results: Collapse to Four Accounts}
\label{sec:collapse}

The central empirical finding is that the leading left eigenvector of a transition matrix spanning 76,855 wallets assigns the overwhelming majority of its mass to exactly four wallets, representing just 0.01\% of the sample, with non-normalized eigenvector values of 0.577, 0.418, 0.359, and 0.305. These are several orders of magnitude above the remaining wallets
($10^{-2}$ down to $10^{-5}$). Under the absorbing-state interpretation of Theorem~\ref{thm:steady-state}, these four wallets are the long-run destinations toward which aggregate fund flow originating from miner wallets converges.

\begin{table}[h]
\end{table}

Notably, none of the three dominant revenue miners from
Section~\ref{sec:revenue-concentration} (Ethermine, SparkPool, F2Pool) appear among these four terminal accounts: revenue concentration and fund-flow concentration are distinct phenomena, pointing to entirely different wallets, and our methodology detects the latter, which is invisible to standard mining-revenue analysis. The three-tier structure predicted by
the simulation in Section~4 is also confirmed empirically: a dominant tier (4 wallets, $\pi_j \approx 0.3$--$0.6$), an intermediate tier ($\pi_j \approx 10^{-2}$--$10^{-1}$), and a large transient tier (remaining 76,840+ wallets, $\pi_j < 10^{-4}$). 
 
\subsection{Robustness: Second-Order Chain}
\label{sec:robustness}
 Applying the second-order robustness check of
Section~\ref{sec:robustness-extensions} to the miner-wallet dataset, the four dominant terminal wallets retain identical ranks under the second-order marginal $\tilde\pi$, with Spearman rank correlation $\rho_S = 1.000$ across all wallets with non-trivial steady-state mass and maximum pointwise difference $|\pi_j - \tilde\pi_j| < 2 \times 10^{-3}$.
This mirrors the toy-example result of Section~4 and supports the
first-order Markov approximation (Assumption~3.2): conditioning on the prior hop does not materially change the predicted terminal destination in this dataset.

\section{Conclusion}

This paper makes two contributions. The first is methodological: we develop a spectral approach to near-real-time measurement of fund-flow concentration, grounded in the Perron--Frobenius theorem and computed via a single eigenvector decomposition rather than iterative matrix multiplication. Unlike existing decentralization metrics, which are almost entirely origination-side (hashrate share, stake share, node counts), our method measures concentration on the \emph{destination} side: not who
initiates transactions, but where the value they carry ultimately comes to rest. The second contribution is empirical: applied to a transaction graph of 76,855 Ethereum wallets built from block-level mining data spanning January 2019 through August 2023, the methodology reveals that fund flows across this large and ostensibly diverse validator population collapse overwhelmingly to just four terminal accounts.

This finding has a double structure we regard as its most notable
feature. Mining revenue on Ethereum is already concentrated by
conventional measures (Section~\ref{sec:revenue-concentration}): the top 10\% of wallets generate 5{,}706$\times$ the ETH revenue of the bottom 90\%, and the top three wallets by revenue outearned the remaining thousands combined. Yet none of these three dominant revenue miners (Ethermine, SparkPool, F2Pool) are among the four terminal-state accounts identified by the eigenvector. The origination-side concentration (who
earns the rewards) and destination-side concentration (where the value ends up) are distinct phenomena that, in this dataset, point to entirely different wallets. A network only moderately concentrated by revenue share can nonetheless direct nearly all of its long-run fund flow to a handful of terminal destinations, a sharper form of centralization than block-reward statistics alone would suggest and one invisible to the metrics the existing decentralization literature relies on (Section~2). This is reinforced by the regression evidence of Section~\ref{sec:revenue-concentration}: top-decile miners show a
significant relationship between block quality and revenue that is absent in lower deciles, consistent with selective block capture. If a subset of validators can selectively construct the blocks they mine, downstream fund-flow concentration may reflect a more structured, non-random routing process than either a naive random-walk model or origination-side revenue statistics alone would reveal.

This concentration also has secondary relevance for illicit-finance monitoring: three of the four terminal accounts we identify are independently associated, post hoc, with documented network irregularities, though we emphasize that our methodology detects structural concentration of fund flow, not intent, and that legitimate high-volume destinations would produce an identical signature.

\paragraph{Limitations and future directions.} Two limitations bound the claims made here. First, the absence of labeled ground truth means that true and false positive rates for the terminal-score and closure statistics remain unknown; validation against labeled datasets such as Elliptic~\cite{weber2019aml} or OFAC-sanctioned address lists is the most important direction for future work and a prerequisite for enforcement-grade conclusions. Second, our results reflect a single terminal-account snapshot from miner-adjacent wallets over a four-year
window; establishing whether the same four accounts persist across Ethereum's distinct validation regimes (POW, Beacon, POS) despite near-complete turnover in the underlying mining population is the most immediate next step, since such persistence would constitute strong evidence that destination-side concentration is structural rather than an artifact of this sample. Absent that test, and absent identification of each terminal account's controlling entity~\cite{trozze2023degens}, our findings should be read as a measurement of fund-flow structure rather than a conclusion about any specific wallet's legitimacy.

Our findings suggest that the apparent diversity of the Ethereum miner network masks a much simpler underlying topology at the level of fund flows: a small number of terminal accounts toward which value generated by a large, seemingly decentralized validator population reliably converges. Whether this reflects benign structural features, such as the routine dominance of
centralized exchanges as cash-out venues, or a more consequential erosion of blockchain's decentralization promise, we leave as an open empirical question. This destination-side methodology is designed to help answer it.

\bibliography{AI,References,KYC_fromEC,FundFlowsBlockchains, fraud}

\end{document}